\documentclass[12pt]{article}

\usepackage{latexsym}  
\usepackage{amssymb}
\usepackage{amsmath}

\def\io {\overline{\imath}}
\def\jo {\overline{\jmath}}
\def\zo {\overline{z}}

\def\psio {\overline{\psi}}

\def\pz  {\partial_z}
\def\pzo {\partial_{\overline{z}}}
\def\eib {\;e^{i\beta}\;}
\def\eibb{\;e^{-i\beta}\;}
\def\pt  {\; \partial_t}
\def\to  {\overline{\theta}}
\def\Ao  {\overline{A}}

\begin{document}
\noindent             

\begin{titlepage}                   

\title{\Large \bf Integrability of the   ${\cal N}=2$   boundary sine-Gordon  model}  

\author{{\Large Tako Mattik}\thanks{mattik@phys.ethz.ch}\\[4ex]   Institut f\"ur Theoretische Physik
\\ ETH Z\"urich\\CH-8093 Z\"urich\\ Switzerland}
\date{21.01.06}
\maketitle \abstract{We construct a boundary Lagrangian for  the ${\cal N}=2$ supersymmetric  sine-Gordon model which preserves (B-type)  supersymmetry and integrability  to all orders in the bulk coupling constant $g$. The supersymmetry constraint is expressed in terms of matrix factorisations.}

\thispagestyle{empty}
\end{titlepage}

\section{Introduction}\label{intro}
Soon after the work of Ghoshal and Zamolodchikov in \cite{ghoshal} on factorisable S-matrices and boundary states in integrable  boundary models, Warner  considered in \cite{warner1} boundary  theories which possess  in addition   a  ${\cal N}=2$ supersymmetry structure.  The key point of his  approach was to include fermionic fields solely defined on the boundaries as previously  introduced in \cite{ghoshal} to ensure unbroken supersymmetries in the boundary theory.\\  In a  similar spirit,  Nepomechie thereafter considered  supersymmetric extensions of the sine-Gordon model. In \cite{nepo1} and \cite{nepo2}  he constructed  boundary Lagrangians  for the ${\cal N}=1$ and ${\cal N}=2$  cases,  establishing  supersymmetry {\em and} integrability which beforehand were  considered to be incompatible in the presence of boundaries. For a nice review of this development see \cite{nepo3}  and \cite{nepo4,baseilhac} for further related  results.\\ The authors of \cite{baseilhac} especially  provide an explanation for  the appearance of fermionic  boundary  degrees of freedom following \cite{baseilhac2, baseilhac3} by using a perturbative CFT approach. See also \cite{baseilhac4} for a classification of admissible  boundary conditions in this context. \\ Whereas the Lagrangian for the ${\cal N}=1$ case in the treatment  of  \cite{nepo1} is exact, the Lagrangian from \cite{nepo2} ensures integrability only to first order in the bulk coupling constant $g$. Up to this order the boundary Lagrangian contains three free continuous  parameters.\\[1ex] The main goal of this paper is to extend the discussion of \cite{nepo2} and to construct a boundary Lagrangian which preserves  supersymmetry and integrability in the sense of a conserved higher spin quantity as in \cite{nepo2} to {\em all} orders in the bulk coupling constant. Our final result will contain up to phases and discrete choices only a single  free boundary parameter. The additional parameters from \cite{nepo2} are fixed in our case by constraints of higher order in the bulk coupling constant.\\ While our main focus will be the (Lorentzian) ${\cal N}=2$ sine-Gordon theory, the supersymmetry considerations in  the first part are valid for arbitrary superpotentials.\\[2ex]
One of our initial motivations to consider the present  problem is the appearance of the sine-Gordon model as the worldsheet theory describing strings in a particular Maldacena-Maoz background from \cite{maldacena}. These backgrounds are particular pp-wave solutions  preserving  at least 4 space-time supersymmetries. For a flat transverse space they are exact superstring solutions \cite{berkovits, russo} whose worldsheet theories  in  light-cone gauge are given by ${\cal N}=(2,2)$ supersymmetric Landau-Ginzburg models \cite{maldacena}.\\  Branes in these general backgrounds  without the inclusion of boundary fields have been studied in \cite{hikida}. Our results give rise to  additional supersymmetric brane configurations. It would be  interesting to obtain a more  detailed understanding of these  new  branes, in particular  from a space-time point of view.\\[2ex] Boundary Lagrangians with fermionic excitations   as used in \cite{warner1, nepo2}  appeared recently in a different  string theoretical context  in \cite{kapustin, brunner1}. These  authors gave a realisation of a suggestion by  Kontsevich to characterise B-type branes in particular Landau-Ginzburg models related to superconformal field theories in terms of matrix factorisations.  This proposal was further studied in \cite{kapustin2, lazar, herbst, brunner2, walcher1, brunner3, brunner4, hori2,  enger}, see also \cite{hori} for a review and further references.\\ The boundary Lagrangian we will use for the sine-Gordon model is of the same type as  those used in \cite{kapustin, brunner1}. It would be very interesting to see how to  employ the methods of the previous references in the case of non-homogeneous worldsheet superpotentials to obtain further insight  for example  about the spectrum of  the boundary theory.\\[2ex]
 Motivated by  possible applications  in string theory we  define the  boundary theories in this paper  on a strip with topology $\mathbb{R}\times [0,\pi]$  instead of the half space as  in \cite{nepo2}. The differences have only notational significance and do not affect any conclusions.\\ A possibly  more serious difference compared to \cite{nepo2} derives from our choice of a Lorentzian worldsheet signature compared to the Euclidean setting in \cite{nepo2}. The structure of the bosonic fields is  almost unaffected  by this.  Deviating reality properties  of the fermions, however, make a direct comparison of the fermionic sectors subtle and we do not attempt to relate them via a Wick rotation.\\ Although the indefinite worldsheet metric  does not directly  simplify the calculations concerning the integrability, its consequences in particular in the fermionic sector make reality requirements more transparent than in \cite{nepo2}. This will be  especially  helpful for  studying the structure of  the  boundary potential $B(z,\overline{z})$ and the conserved supersymmetries.\\[2ex]
The  paper is organised as follows. In the subsequent  section \ref{blagra} we will write down a  boundary Lagrangian including  fermionic boundary excitations and derive the resulting boundary conditions. In section \ref{boundconserv} we will find  conditions  under  which the boundary theory is ${\cal N}=2$ supersymmetric. The relevant requirement  will be the matrix factorisation constraint  from  \cite{kapustin, brunner1}.\\ In the final section \ref{intesin} the integrability of  the  boundary theory for  the particular  sine-Gordon case will be considered. Additional information such as  the explicit component form of the higher spin conserved currents  from \cite{kobayashi, nepo2}, are supplied in the appendix.
  
\section{Landau-Ginzburg models and boundary fermions}\label{blagra}
In the first part we will consider general   ${\cal N}=2$ supersymmetric Landau-Ginzburg models  with flat target space and vanishing holomorphic Killing vector term.
 On worldsheets without boundaries these theories are  described by the (component)  Lagrangian\begin{eqnarray} \nonumber  {\cal L}_{\mbox{bulk}}  &=& \frac{1}{2} g_{j\jo} \left(\partial_+ z^j \partial_- \zo^{\jo}+\partial_+ \zo^{\jo}\partial_- z^j+i \overline{\psi}^{\jo}_+ \stackrel{\leftrightarrow}{\partial}_- \psi^j_++i\overline{\psi}_-^{\jo}\stackrel{\leftrightarrow}{\partial}_+ \psi^j_-  \right)\\ &\;&\; -\frac{1}{2}\partial_i\partial_j W({\bf z})\;  \psi^i_+\psi^j_- - \frac{1}{2}\partial_{\io}\partial_{\jo}\overline{W}(\overline{\bf z})\; \overline{\psi}^{\io}_-\overline{\psi}_+^{\jo}-\frac{1}{4}g^{i\jo} \partial_iW({\bf z})\; \partial_{\jo}\overline{W}(\overline{\bf z}), \label{lagragen}    \end{eqnarray} with $\partial_\pm=\pt\pm \partial_\sigma$,  compare for example with \cite{vafa1, vafa2, vafa3}.\\ When defined on a manifold with boundaries, one might either enforce boundary conditions in addition to  the equations of motion from (\ref{lagragen}) or include   boundary terms  containing in particular   fermionic boundary degrees of freedom and work with the resulting (boundary) equations of motion.  To the best of our knowledge, the second approach goes back to the   study of  integrals of motion and factorisable S-matrices of  integrable boundary theories in the  seminal paper   \cite{ghoshal}.  There the  authors  considered in particular the massive Ising model with boundary fermions  and the bosonic sine-Gordon theory with an additional bosonic boundary potential.\\  Comparable boundary Lagrangians  have   thereafter been adopted in different ways in  \cite{warner1} and  \cite{nepo1, nepo2}  in the context of supersymmetric integrable boundary field theories. For a construction of the boundary structure  relying on a boundary quantum group  see  \cite{baseilhac, baseilhac2}.\\  As mentioned in the introduction, the approach following \cite{warner1, nepo2}  has also recently appeared in a string theory context, compare for example with  \cite{kapustin, brunner1}.\\[2ex]
In this paper we  use a  boundary Lagrangian following \cite{nepo2}, see also \cite{kapustin, brunner1},  given by 
\begin{eqnarray}\nonumber {\cal L}^{\sigma=\pi}_{\mbox{boundary}} &=&\frac{i}{2}\left(b\psi_-\psio_+   -b^* \psi_+\psio_- \right) -\frac{i}{2}a\stackrel{\leftrightarrow}{\partial}_t\overline{a}+B(z,\zo)  \\  \nonumber &\;\;& +\frac{i}{2} \left(\overline{F}'(\zo)\overline{a}+\overline{G}'(\zo)a\right)\left(\psio_++e^{i\beta}\psio_-\right)\\  &\;\;&+\frac{i}{2}\left(\rule{0mm}{4mm}
G'(z)\overline{a}+F'(z)a\right)\left(\psi_++e^{-i\beta}\psi_-\right).\label{boundarylagra}   \end{eqnarray}
As we will later on restrict attention to superpotentials depending on a single holomorphic coordinate, the boundary Lagrangian   is  written down containing only contributions along  the $z=z^1$  direction at $\sigma = \pi$. The Lagrangian (\ref{boundarylagra})  is chosen to be manifestly real and the constant $b$  is determined by consistency of the resulting (fermionic) boundary conditions to  $b=e^{-i\beta}$, compare for example with \cite{ameduri}.
\subsection{The boundary conditions}
In this part  we will determine    the boundary conditions resulting from the boundary Lagrangian  (\ref{boundarylagra}). They  are  obtained from the variation of ${\cal L}_{\mbox{boundary}}$ together with  boundary terms  from partially integrated contributions  in $\delta{\cal L}_{\mbox{bulk}}$.
\\ The bosonic part of the bulk contributions from (\ref{lagragen})  is  given by 
 \begin{eqnarray} \left. -g_{i\io}\left(\delta z^i \partial_\sigma \overline{z}^{\io}+\delta \overline{z}^{\io}\partial_\sigma z^i  \right)\right|_{\sigma=0}^\pi = \left. -2\delta x^I \partial_\sigma x^I\right|_{\sigma=0}^\pi,\end{eqnarray}
whereas the fermionic kinetic parts lead  to 
 \begin{eqnarray}\left. \frac{1}{2} g_{i\io}\left(-i\psio^{\io}_+\delta\psi^i_++i\delta\psio^{\io}_+ \psi_+^i +i \psio^{\io}_-\delta\psi^i_--i\delta\psio^{\io}_-\psi^i_-  \right) \right|_{\sigma=0}^\pi  =i\left.\left( \psi_-^I \delta\psi_-^I-\psi^I_+\delta\psi_+^I \right)\right|_{\sigma=0}^\pi  .\end{eqnarray}Altogether, the   boundary conditions for the $z=z^1$ direction  at $\sigma=\pi$  are with these terms found to be  \begin{eqnarray} \partial_\sigma z &=& \partial_{\zo}B(z,\zo)+\frac{i}{2}\left(\overline{F}''(\zo)\overline{a}+\overline{G}''(\zo)a\right)\left(\psio_++e^{i\beta}\psio_-\right)\\ \partial_t a &=& \frac{1}{2}\overline{F}'(\zo)\left(\psio_++e^{i\beta}\psio_-\right)+\frac{1}{2}G'(z)\left(\psi_++e^{-i\beta}\psi_-\right)\\ \psi_+-e^{-i\beta}\psi_- &=& \overline{F}'(\zo)\overline{a}+\overline{G}'(\zo)a   \end{eqnarray}together with  the complex conjugates \begin{eqnarray} \partial_\sigma \zo &=& \partial_z B(z,\zo)+\frac{i}{2}\left(G''(z)\overline{a}+F''(z)a\right)\left(
\psi_++e^{-i\beta}\psi_-\right)\\ \partial_t\overline{a} &=& \frac{1}{2}\overline{G}'(\zo)\left(\psio_++e^{i\beta}\psio_-\right)+\frac{1}{2}F'(z)\left(\psi_++e^{-i\beta}\psi_-\right)\\ \psio_+-e^{i\beta}\psio_- &=& G'(z) \overline{a}+F'(z)a.\end{eqnarray} \\
Setting \begin{eqnarray}  \label{aofz} A(z) &=& G'(z)\overline{a}+F'(z)a\\  \label{abarofz}\overline{A}(z)&=& \overline{F}'(\zo)\overline{a}+\overline{G}'(\zo)a \end{eqnarray} and using the suitable  fermionic combinations \begin{eqnarray} \nonumber \theta_+= \frac{1}{2} \left(\psi_++e^{-i\beta}\psi_-\right)\;\;\; &\;&\;\;\; \overline{\theta}_+=\frac{1}{2} \left(\overline{\psi}_++e^{i\beta}\overline{\psi}_-\right)\\  \theta_-= \frac{1}{2} \left(\psi_+-e^{-i\beta}\psi_-\right)\;\;\; &\;&\;\;\; \overline{\theta}_-=\frac{1}{2} \left(\overline{\psi}_+-e^{i\beta}\overline{\psi}_-\right)  \end{eqnarray}
\begin{eqnarray}\nonumber \psi_+ =\theta_+ + \theta_- \;\;\; &\;& \;\;\; \psi_- = e^{i\beta} \left(\theta_+-\theta_-\right)\\  \overline{\psi}_+= \overline{\theta}_++\overline{\theta}_-    \;\;\; &\;& \;\;\; \overline{\psi}_- = e^{-i\beta}\left(\overline{\theta}_+-\overline{\theta}_-\right)\end{eqnarray}
 the boundary conditions finally become \begin{eqnarray} \label{bc1}\partial_\sigma z  &=& \partial_{\zo}B(z,\zo)+i\overline{A}'(\zo)\overline{\theta}_+\\ \label{bc2} \partial_t a &=& \overline{F}'(\zo)\overline{\theta}_++G'(z)\theta_+\\ \label{bc3}\theta_- &=& \frac{1}{2}\overline{A}(\zo)\end{eqnarray} and \begin{eqnarray}\label{bc4} \partial_\sigma\zo &=&\partial_z B(z,\zo)+iA'(z)\theta_+\\ \label{bc5} \partial_t \overline{a}&=&\overline{G}'\overline{\theta}_++F'(z)\theta_+\\ \label{bc6} \overline{\theta}_- &=&\frac{1}{2} A(z).\end{eqnarray} 
By eliminating the fermionic boundary degrees of freedom  in favour of $\theta_-, \overline{\theta}_-$  in (\ref{bc1}) and (\ref{bc4}),  these bosonic boundary conditions take on the structure  with  a quadratic fermionic correction term  as  already discussed in \cite{lindstrom} from a different point of view.\\
In the next section we will discuss how the so far undetermined holomorphic  functions $F,G$ are related to the superpotential if the boundary theory is to preserve B-type supersymmetries.
\section{Matrix factorisation and ${\cal N}=2$ supersymmetry}
\label{boundconserv}
The Landau-Ginzburg bulk theory of  (\ref{lagragen}) has  the  four  conserved supercurrents \cite{vafa2, vafa3} \begin{eqnarray}\label{susyflux1} G^0_\pm = g_{i\jo}\partial_\pm\zo^{\jo}\psi^i_\pm\mp \frac{i}{2}\psio^{\jo}_{\mp}\partial_{\jo}\overline{W} &\;\;\;\;& G^1_\pm = \mp g_{i\jo}\partial_\pm\zo^{\jo}\psi^i_\pm-\frac{i}{2}\psio^{\jo}_\mp \partial_{\jo}\overline{W} \\ \label{susyflux2} \overline{G}^0_\pm=g_{i\jo}\psio^{\jo}_\pm\partial_\pm z^i\pm\frac{i}{2}\psi^i_\mp\partial_i W  &\;\;\;\;& \overline{G}^1_\pm = \mp g_{i\jo}\psio^{\jo}_\pm\partial_\pm z^i+\frac{i}{2}\psi^i_\mp\partial_i W, \end{eqnarray} whose corresponding charges \begin{equation} Q_{\pm}=\int_0^{2\pi}d\sigma\; G^0_{\pm};\;\;\;\;\; \overline{Q}_{\pm}=\int_0^{2\pi}d\sigma\; \overline{G}^0_{\pm} \end{equation} represent the usual ${\cal N}=(2,2)$ bulk supersymmetry.\\ As usual, the introduction of boundaries breaks  at least a certain number of bulk symmetries. As explained in \cite{vafa2}, there are  essentially  two possibilities to preserve a ${\cal N}=2$ supersymmetry algebra  resulting from  (\ref{susyflux1}) and (\ref{susyflux2}). Here we will  concentrate on the so called B type case.\\
Following \cite{ghoshal, vafa2},  the  (B type) supersymmetries take on    the general  form \begin{eqnarray}\label{qboundary1} Q& =& \overline{Q}_+ + e^{i\beta}\overline{Q}_-+\Sigma_\pi(t)-\Sigma_0(t)\\ \label{qboundary2}  Q^\dagger &=& Q_++e^{-i\beta}Q_-+\overline{\Sigma}_\pi(t)-\overline{\Sigma}_0(t)\end{eqnarray} which includes ({\em local}) contributions of  the boundary fields at $\sigma=\pi$ and $\sigma=0$.  Using the conservation of the bulk fluxes (\ref{susyflux1}) and (\ref{susyflux2}),  the boundary supersymmetries $Q, Q^\dagger$ are time independent, that is, conserved,  if the fluxes fulfil the equations
\begin{eqnarray}\label{fluxw3}  0&=&\left. \overline{G}_+^1+e^{i\beta} \overline{G}^1_-\right|_{\sigma=\pi} -\dot \Sigma_\pi(t)\\  \label{fluxw4}  0&=&\left. \overline{G}_+^1+e^{i\beta} \overline{G}^1_-\right|_{\sigma=0} -\dot \Sigma_0(t) \end{eqnarray} together with their  corresponding complex conjugates. \\ The boundary field $\Sigma_\pi(t)$ ($\Sigma_0(t)$) is here required to depend only on the bulk fields and their time derivatives at  time $t$ evaluated at  $\sigma=\pi$  ($\sigma = 0$) and the boundary degrees of freedom $a(t)$ and $\overline{a}(t)$.   
\subsection{$W$ - factorisation}
In this section we will solve the equation (\ref{fluxw3}) to obtain a condition for  the boundary fields $F(z), G(z)$ and the boundary potential $B(z,\zo)$ for ${\cal N}=2$ supersymmetric branes.\\
From (\ref{fluxw3})  and (\ref{susyflux2}) we obtain   
\begin{eqnarray}  \nonumber \partial_t \Sigma_\pi(t)&\stackrel{!}{=}& \overline{G}^1_++e^{i\beta}\overline{G}^1_-\\  &=& -\psio_+\partial_+z+   e^{i\beta}\psio_-\partial_-z+ \frac{i}{2} \left(\psi_-+e^{i\beta}\psi_+\right)\partial_z W \label{rechnung1}  \end{eqnarray} evaluated at $\sigma = \pi$.   Upon partial integration   (\ref{rechnung1}) leads to
\begin{eqnarray}\nonumber \partial_\tau \Sigma_\pi(t)&=&  -\partial_t z\left(\psio_+-e^{i\beta}\psio_-\right) - \partial_\sigma  z\left(\psio_+ +e^{i\beta}\psio_-\right)\\
&\;\;& + \nonumber \frac{i}{2} \partial_z W(z)\left(\psi_-+e^{i\beta}\psi_+\right)\\
&=& \nonumber  -\partial_t\left(2 z\to_-\right) + z\left(\left(G''(z)\dot z \overline{a}+F''(z)\dot z a \right)+\left(G'(z)\dot{\overline{a}}+F'(z)\dot a\right)\right)\\&\;\;&-2\to_+\partial_{\zo}B(z,\zo) + {i}\eib\theta_+\partial_z W(z)
\end{eqnarray}
from which we get
\begin{eqnarray}\nonumber \partial_t \Sigma_\pi(t)&=&  -\partial_t\left( 2z\to_--p(z)a-q(z)\overline{a}\right)\\ \nonumber
&\;\;&+ \left(zG''(z)-q'(z)\right)\dot z \overline{a}+\left(zF''(z)-p'(z)\right)\dot z a \\  \nonumber &\;\;&+\left(zG'(z)-q(z)\right)\dot{\overline{a}}+\left(zF'(z)-p(z)\right)\dot a\\&\;\;&-2\to_+\partial_{\zo}B(z,\zo) + {i}\eib\theta_+ \partial_z W(z).\end{eqnarray}
Using \begin{equation} q'(z)=zG''(z)\Rightarrow q(z)=zG'(z)-G(z)\end{equation} and \begin{equation} p'(z)=zF''(z)\Rightarrow p(z)=zF'(z)-F(z)\end{equation} together with  the equations of motion for $a$ and $\overline{a}$ we  finally arrive   at
\begin{eqnarray} \nonumber \partial_t \Sigma_\pi(t)&=&  -\partial_t\left( 2z\to_--p(z)a-q(z)\overline{a}\right)\\ \nonumber
&\;\;&+\theta_+\left(G(z)F'(z)+F(z)G'(z)+ie^{i\beta}\partial_z W(z)\right)\\ &\;\;& + 2\to_+\left(\frac{1}{2} G(z)\overline{G}'(\zo)+\frac{1}{2}F(z)\overline{F}'(\zo)-\partial_{\zo}B(z,\zo)\right). \end{eqnarray} The conditions  for ${\cal N}=2$ supersymmetry therefore read \begin{eqnarray}\label{condw}  W(z) &=& ie^{-i\beta}F(z)G(z)+\mbox{const}\\  \label{condb} B(z,\zo)&=& \frac{1}{2} \left(F(z)\overline{F}(\zo)+G(z)\overline{G}(\zo)\right)+\mbox{const}\end{eqnarray}  and the local  boundary field  $\Sigma_\pi$  appearing in the `boundary adjusted'  supercharges (\ref{qboundary1}) and (\ref{qboundary2}) is given by \begin{equation} \Sigma_\pi= -2z\to_-+\left(zF'-F\right)a+\left(zG'-G\right)\overline{a}.\end{equation} It explicitly contains contributions from the fermionic boundary degrees of freedom, compare for example with the results in \cite{baseilhac}.\\[1ex]
The condition (\ref{condw}) is of course identical to the matrix factorisation condition from \cite{kapustin, brunner1}, whereas (\ref{condb}) so far only determines the structure of the boundary potential $B(z,\zo)$.  It does not lead to a   condition on $F,G$ as the boundary potential remains functionally undetermined by the supersymmetry considerations.\\[1ex] In the context of matrix factorisations in string theory as in \cite{kapustin, brunner1} and the literature mentioned in the introduction  the focus is on quasi-homogeneous superpotentials which lead in the infrared to a superconformal field theory. The latter requires a conserved $U(1)$ R-charge which should also be present in the boundary theory, compare for example with \cite{hori2}. This additional condition requires factorisations of $W$ into quasi-homogeneous functions.\\  It is worth pointing out  that there is no corresponding restriction on $F$ and $G$ in our case. Integrability together with supersymmetry in the context of the boundary Lagrangian (\ref{boundarylagra}) will give  particular trigonometric functions in the case of the sine-Gordon model, but supersymmetry on its own allows for more general choices.\\Extending  our treatment, one might following \cite{baseilhac}  add purely bosonic boundary degrees of freedom to  (\ref{boundarylagra}).  This  opens up the possibility for more general choices of $F$ and $G$ even when enforcing supersymmetry and integrability. We will not consider this possibility in this paper.
\section{The ${\cal N}=2$ boundary sine-Gordon model}\label{intesin}
From now on we will specify the   superpotential to   \begin{equation} W(z)=-i\lambda \cos\left(\omega z\right), \label{superpotential} \end{equation} and restrict  attention therefore to the ${\cal N}=2$ supersymmetric sine-Gordon model\footnote{The phase accompanying the real coupling constant $\lambda$ is chosen for later convenience. Its form does not affect purely  bosonic  terms in the Lagrangian (\ref{lagragen}). In the  fermionic parts of (\ref{lagragen})  it can be absorbed in a redefinition $\psi_{\pm}\rightarrow e^{i\alpha}\psi_{\pm}$, $\psio_{\pm}\rightarrow e^{-i\alpha}\psio_{\pm}.$}. \\ When defined on a manifold without boundaries this theory   is well known to be a  supersymmetric and  integrable extension of the purely bosonic sine-Gordon theory~\cite{kobayashi}.  Its first nontrivial conserved higher spin currents on whose conservation in the presence of a boundary we will concentrate in the following,  were derived  in   \cite{kobayashi, nepo2}. In  our conventions they are given  in  the Appendix \ref{cocurrents}.\\
By using (\ref{superpotential}) in (\ref{lagragen}), one can immediately derive the bulk equations of motion. They are given by
\begin{eqnarray}\label{eom1} \partial_+\partial_- z &=& -ig\sin\zo\;\; \psio_-\psio_+-g^2\sin z\cos\zo\\ \label{eom2} \partial_+\partial_- \zo &=& ig \sin z\;\; \psi_+\psi_- -g^2\cos z\sin \zo\\ \label{eom3}  \partial_-\psi_+ &=& g \cos\zo\;\; \psio_-\\ \label{eom4} \partial_-\psio_+ &=& g \cos z\;\; \psi_-\\ \label{eom5} \partial_+\psi_- &=& -g \cos\zo\;\; \psio_+\\  \label{eom6} \partial_+\psio_- &=& -g\cos z\;\; \psi_+,\end{eqnarray} where we set  $\omega =1$ and redefined the bulk coupling constant to $g=\frac{\lambda}{2}$, resembling the choices in \cite{nepo2}. 
\subsection{Integrability in the presence of a boundary}
In this section we will consider the ${\cal N}=2$ sine-Gordon model in the presence of a boundary and derive conditions under which the following `energy-like' combination of the bulk conserved quantities  \begin{equation}\label{spin3b} I_3=\int_0^\pi d\sigma\; \left(T_4+\overline{T}_4-\theta_2-\to_2\right) -\Sigma_\pi^{(3)}(t)+\Sigma_0^{(3)}(t) \end{equation}  is conserved when using the boundary  Lagrangian (\ref{boundarylagra}). The       inclusion of local boundary currents  as  $\Sigma^{(3)}_0(t)$ and $\Sigma^{(3)}_\pi(t)$  goes back to  \cite{ghoshal}. Their  appearance  is by now a well known and frequently used  feature in the context of integrable boundary field theories.   It  is in particular independent of the in our case  present  supersymmetries. \\   The conservation of a higher spin  quantity like $I_3$ is usually regarded as providing  strong evidence for  the  integrability of the underlying two dimensional (boundary)  field theory.\\[1ex]  
As previously done in  section  \ref{boundconserv}  for the supercurrents (\ref{susyflux1}), (\ref{susyflux2}) in  (\ref{qboundary1}) and (\ref{qboundary2}),  the quantity $I_3$ is conserved  if the condition \begin{equation} \label{cond3b}\partial_t \Sigma_\pi^{(3)} =  T_4-\overline{T}_4+\theta_2-\to_2\end{equation} holds at $\sigma=\pi$. In deriving (\ref{cond3b}) we have used the equations (\ref{conservi1}) and (\ref{conservi2}) from the appendix \ref{cocurrents}.  As before, there is an  identical equation at $\sigma = 0$.\\[2ex] Due to the complexity of the conserved currents as given in the appendix \ref{cocurrents},  the calculation transforming the right hand side of  (\ref{cond3b})  to a total time derivative is rather lengthy and intricate. It nevertheless follows a  straightforward strategy which in our case differs slightly from the approach in \cite{nepo2}.\\[1ex] In a first step we use the equations of motion (\ref{eom1})-(\ref{eom6}) and the bosonic boundary conditions (\ref{bc1}) and (\ref{bc4})  to remove all  $\sigma$ derivatives on the bosonic and  fermionic fields appearing in  $T_4,\overline{T}_4,\theta_2,\to_2$.\\ In a second step we  remove (where possible) all time derivatives on the fermionic fields $\theta_+$ and $\to_+$ by partial integration  and apply the identities from appendix  \ref{aidentity}  to furthermore replace $\theta_-$ and $\to_-$ and their time derivatives by the fermionic boundary fields  $a$ and $ \overline{a}$. \\ In doing so  a large number of terms cancel manifestly. There are,  however,  other terms  as for example  those proportional to combinations like $\left(\theta_+\pt \theta_+\right)$ or $\left(\theta_+\to_+\right)$ which cannot be reduced further and which  cannot be written as a time derivative of a local field. Their prefactors given by expressions containing    the boundary potential $B(z,\zo)$ and the functions $F(z), G(z)$ and their derivatives   therefore necessarily  have to vanish. \\ Together with the conditions (\ref{condb}) and (\ref{condw}) for  the ${\cal N}=2$ supersymmetry these resulting differential equations actually will be seen to  determine the boundary Lagrangian up to two possible choices for the boundary potential including a free parameter and two additional (discrete) choices in  prefactors appearing in  the functions $F$ and $G$. 
\\[2ex] In the following we will write down the differential equations determined as explained above and give  their solutions. The explicit form of the boundary field  $\Sigma_\pi^{(3)}$ appearing in (\ref{spin3b})  will be provided  in the appendix.

\subsection{The boundary potential $B(z,\zo)$} 
As   explained in \cite{nepo2}, the boundary potential $B(z,\zo)$ is already determined from the purely bosonic  terms (\ref{t1}), (\ref{tb1}) and (\ref{o5}), (\ref{ob5}).  The differential equations for the {\em real} field $B$ read  \begin{eqnarray} 0&=& \pz\pz\pzo B+\frac{1}{4} \pzo B\\ 0&=& \pzo\pzo\pz B + \frac{1}{4}\pz B \end{eqnarray} together with \begin{eqnarray}\pz\pz B =\pzo\pzo B.\end{eqnarray} This determines $B$ to \begin{eqnarray}\label{boundpot}  B(z,\zo) = \alpha \cos\frac{z-z_0}{2}\cos\frac{\zo-\zo_0}{2}+b,\;\;\; \alpha,b\in\mathbb{R}, z_0\in \mathbb{C}\end{eqnarray} which  is so far exactly the result of \cite{nepo2}. Together with (\ref{condb}) we will nevertheless find further conditions on the so far unspecified constant  $z_0$ which come from contributions of  higher  order in the bulk coupling constant $g$ than considered in \cite{nepo2}.
\subsection{The boundary functions  $F, G,$ and $\overline{F}, \overline{G}$}
From terms quadratic in the fermionic degrees of freedom
as for example from \begin{eqnarray}16i\left(\pt z\right)^3 \left(A'''(z)+\frac{1}{4}A'(z)\right)\theta_+ \end{eqnarray} and \begin{equation} 48 i \pt z\; \pt^2 z\; \left(A''(z)+\frac{1}{4}A\right)\theta_+\end{equation}
 we obtain  the differential equations \begin{eqnarray}\label{adoppel}  0&=& A''(z)+\frac{1}{4}A(z) \\ \label{adoppel2} 0 &=&\pz\left[F'(z)G'(z)\right]+\frac{1}{2}g\sin z\; \eib \end{eqnarray} together with  the corresponding complex conjugates.\\ From (\ref{adoppel}) and (\ref{aofz}), (\ref{abarofz}) the functions  $F(z)$ and $G(z)$ are determined to \begin{eqnarray}\label{F1} F(z) &=& A_0 \cos \frac{z-\kappa_1}{2}+C_0\\ \label{G1} G(z) &=& B_0 \cos\frac{z-\kappa_2}{2}+D_0,\end{eqnarray} and the equation (\ref{adoppel2}) becomes  with the matrix factorisation condition (\ref{condw}) \begin{eqnarray}\nonumber 0 &=& F'\left(G''+\frac{1}{4}G\right)+G'\left(F''+\frac{1}{4}F\right)\\ &=& -\frac{1}{8}\left( A_0D_0 \sin\frac{z-\kappa_1}{2}+B_0C_0\sin\frac{z-\kappa_2}{2}\right). \end{eqnarray} 
 By combining these results with the expression for $B(z,\zo)$ found in (\ref{condb}), we can in the next step deduce conditions on the so far free parameters in (\ref{F1}), (\ref{G1}) and (\ref{boundpot}).\\[2ex] Using (\ref{F1}) and (\ref{G1}) in the differentiated condition (\ref{condw}) 
we obtain \begin{equation} 2g\eib \sin z = -\frac{A_0B_0}{2}\cos\frac{\kappa_1+\kappa_2}{2}\sin z+ \frac{A_0B_0}{2}\sin\frac{\kappa_1+\kappa_2}{2}\cos z\end{equation} and therefore \begin{eqnarray}\label{ab}  \kappa_1+\kappa_2 &=& 2\pi n\;\;\;n\in \mathbb{Z}\\ 2g\eib &=&-\frac{1}{2} A_0B_0\; (-)^n.\end{eqnarray} A constraint  on $z_0$ appearing in the boundary potential $B$ comes from equation (\ref{condb}). In particular, we have \begin{equation} \label{condb2} \pz\pzo B(z, \zo) = \frac{1}{2} \left( F' \overline{F}'+G'\overline{G}'\right).\end{equation} Using (\ref{boundpot}) and (\ref{F1}), (\ref{G1}) evaluated at $z= z_0$ we get \begin{equation} 0=A_0\overline{A}_0\sin\frac{z_0-\kappa_1}{2}\sin\frac{\zo_0-\overline{\kappa_1}}{2}+B_0\overline{B}_0\sin\frac{z_0-\kappa_2}{2}\sin\frac{\zo_0-\overline{\kappa_2}}{2}\end{equation} and  therefore  \begin{equation} 0= \sin\frac{z_0-\kappa_1}{2};\;\;\; 0=\sin\frac{z_0-\kappa_2}{2}.\end{equation} Together with (\ref{ab}) and the observation that the boundary Lagrangian (\ref{boundarylagra}) does not depend on the constants $C_0, D_0$ in (\ref{F1}) and (\ref{G1}) we therefore obtain the two following possibilities ensuring integrability in the sense discussed above. {}
\paragraph{Case I:} \begin{eqnarray} B(z,\zo) =\alpha \cos\frac{z}{2}\cos\frac{\zo}{2};\;\;   F(z)=A_0\cos\frac{z}{2};\;\; G(z)=B_0\cos\frac{z}{2}\end{eqnarray} with \begin{eqnarray}\label{zwischen2} A_0B_0=-4g\eib;\;\;A_0\overline{A}_0+B_0\overline{B}_0=2\alpha\end{eqnarray} {}  \paragraph{ Case II:}\begin{eqnarray} B(z,\zo)=\alpha \sin\frac{z}{2}\sin\frac{\zo}{2};\;\; F(z)=A_0\sin\frac{z}{2};\;\; G(z)=B_0\sin\frac{z}{2}\end{eqnarray} with \begin{eqnarray} \label{zwischen3}A_0B_0=4g\eib;\;\; A_0\overline{A}_0+B_0\overline{B}_0=2\alpha.\end{eqnarray}
From (\ref{zwischen2}) and (\ref{zwischen3}) we have  in both cases \begin{eqnarray} \label{zwischen4} A_0\overline{A}_0=\alpha\pm\sqrt{\alpha^2-16g^2}\end{eqnarray} and therefore \begin{eqnarray} \label{apm} A_0^\pm=e^{i\gamma} \sqrt{\alpha\pm\sqrt{\alpha^2-16g^2}}.\end{eqnarray}  The undetermined  phase $\gamma$ appearing in (\ref{zwischen4}) can be absorbed in a redefinition of the fermionic boundary fields $a$ and $\overline{a}$.   From  (\ref{zwischen4}) we furthermore have the condition  \begin{equation} \alpha \ge 4g\ge 0\end{equation} which in particular leads to a positive - semidefinite boundary potential $B(z,\zo)$ in (\ref{boundarylagra}).\\[2ex]  With these choices all remaining terms in (\ref{cond3b}) either vanish or can be written as a total time derivative as a rather long calculation shows. This ensures  therefore the conservation of the  higher spin quantity (\ref{spin3b}) in the presence of a  boundary to all orders  in the bulk coupling constant $g$, providing  strong evidence for integrability. \\[1ex]   To first order in $g$,  the condition (\ref{condb2}) together with (\ref{zwischen3}) and (\ref{zwischen4})  does not give a constraint  on $F$ and $G$ and one reobtains the situation of \cite{nepo2} where the two additional (real) parameters expressed by  $z_0$ of the boundary potential  in (\ref{boundpot})  were found to be compatible with integrability to that order.

\section{Conclusions} Following \cite{warner1, nepo2} and \cite{kapustin, brunner1},    we constructed  boundary Lagrangians which establish  supersymmetry and integrability  in the sense of a conserved higher spin current for the (Lorentzian)  ${\cal N}=2$ supersymmetric sine-Gordon model defined on the  strip or the half space.\\ In contradistinction  to \cite{nepo2}, our  Lagrangians are  exact to all orders in the bulk coupling constant $g$. Apart from   phases and the bulk coupling constant  $g$, both possible choices contain a single continuous parameter $\alpha$.  The  additional free complex  parameter $z_0$ appearing in \cite{nepo2} is in our case  essentially fixed to  $z_0=0$ or  $z_0=\pi$.\\[2ex] The  Ansatz (\ref{boundarylagra}) for the boundary Lagrangian secures  supersymmetry for a general superpotential if the boundary functions $F,G$ fulfil the matrix factorisation condition (\ref{condw}). Following \cite{baseilhac} one might furthermore add purely bosonic boundary degrees of freedom to (\ref{boundarylagra}). It would be interesting to study this situation  in greater detail.\\ As mentioned in the introduction,  one can  readily  apply the present setup  to the construction of  branes in Maldacena-Maoz backgrounds \cite{maldacena} leading to branes that generalise the constructions of  \cite{hikida}.  This will be considered elsewhere.\\[1ex] Apart from  the construction of further branes in Maldacena-Maoz backgrounds it would be interesting to obtain  a more detailed (space-time) interpretation of the boundary fermions and the corresponding branes. At least for the integrable  sine-Gordon case it might furthermore be possible to derive detailed information about the  string spectrum by  using methods like  the   thermodynamic Bethe Ansatz, see  for example  \cite{clair, moriconi, bazhanov, dorey, dorey2} and \cite{tirziu} for a  related setting. 
  
\section*{Acknowledgements}
I am  grateful  to  Matthias Gaberdiel
for encouragement and support. I furthermore  thank   Stefan
Fredenhagen and Peter Kaste  for discussions and helpful comments on the manuscript.  This
research is partially  supported by the Swiss National Science
Foundation and the European Network `ForcesUniverse' (MRTN-CT-2004-005104).

\begin{appendix}
\section{Higher spin conserved currents}\label{cocurrents}
Using  the superfield approach from \cite{kobayashi}, the first higher spin conserved bulk currents for the ${\cal N}=2$ supersymmetric sine-Gordon model were determined in components by Nepomechie in \cite{nepo2} to (in our conventions)
\begin{eqnarray}\nonumber \label{t1}  T_4&=& -\left(\partial_+\zo\right)^3\partial_+z-\left(\partial_+z\right)^3\partial_+\zo  +2 \partial_+^2\zo\; \partial_+^2z   \\ \nonumber  &\;\;& -i\psio_+\partial_+\psi_+\left[(\partial_+\zo)^2+3(\partial_+ z)^2\right] -2i \psi_+\partial_+\psio_+ (\partial_+\zo)^2  \\ &\;\;& -2i \psio_+\psi_+\;\; \partial_+\zo \partial_+^2\zo \label{t6}  +2i \partial_+\psio_+ \partial_+^2 \psi_+\\ \nonumber  \theta_2 &=& g^2\sin z \sin\zo \left[(\partial_+\zo)^2+(\partial_+ z)^2\right] -2g^2\cos z \cos \zo\; \partial_+ z \partial_+\zo\\  \nonumber &\;\;& -ig \cos z \psi_-\psi_+ \left[(\partial_+\zo)^2+(\partial_+ z)^2\right]\\  &\;\;&-2ig \sin z \partial_+z \left[-\psi_+\partial_+\psi_-+\psi_-\partial_+\psi_+\right] +2ig \cos z \partial_+ \psi_- \partial_+\psi_+ \nonumber\\ \nonumber  \label{o1}   &=& g^2\sin z \sin\zo \left[(\partial_+\zo)^2+(\partial_+ z)^2\right] -2g^2\cos z \cos \zo\; \partial_+ z \partial_+\zo\\ \nonumber \label{o3} &\;\;& -2ig^2 \sin z \cos\zo \partial_+ z\;\; \psi_+\psio_+  -2ig^2 \cos z \cos \zo \;\; \psio_+\partial_+\psi_+\\ \label{o5}&\;\;& -ig \cos z \psi_-\psi_+ \left[(\partial_+\zo)^2+(\partial_+ z)^2\right] -2ig \sin z \partial_+ z\;\; \psi_- \partial_+\psi_+
\end{eqnarray}
and 
\begin{eqnarray} \nonumber \label{tb1} \overline{T}_4 &=& -(\partial_- \zo)^3 \partial_- z - (\partial_- z)^3 \partial_- \zo  +2 \partial_-^2 \zo \partial_-^2 z  \\  \nonumber \label{tb2} &\;\;& -i \psio_- \partial_- \psi_- \left[ (\partial_- \zo)^2 +3 (\partial_- z)^2\right]  -2i \psi_- \partial_- \psio_- (\partial_- \zo)^2 \\  &\;\;&  -2i \psio_- \psi_- \partial_- \zo \partial_-^2 \zo   \label{tb6}  +2i \partial_- \psio_- \partial_-^2 \psi_-\\ \nonumber \overline{\theta}_2 &=& g^2 \sin z \sin \zo  \left[ (\partial_- \zo)^2 + (\partial_- z)^2\right] -2g^2 \cos z \cos \zo \partial_- z \partial_- \zo \\ \nonumber &\;\;& -ig \cos z \psi_- \psi_+  \left[ (\partial_- \zo)^2 + (\partial_- z)^2\right]\\ \nonumber  &\;\;& -2ig \sin z \partial_- z\; (-\psi_+\partial_- \psi_-+\psi_- \partial_- \psi_+) +2ig \cos z \partial_- \psi_- \partial_- \psi_+ \\ \nonumber  &=& g^2 \sin z \sin \zo  \left[ (\partial_- \zo)^2 + (\partial_- z)^2\right]  -2g^2 \cos z \cos \zo \partial_- z \partial_- \zo \\ \nonumber  &\;\;&-2ig^2 \sin z \cos \zo \partial_- z \psi_- \psio_- +2ig^2 \cos z \cos \zo \partial_-\psi_- \psio_-\\ \label{ob5}&\;\;& -ig \cos z \; \psi_-\psi_+ \left[ (\partial_- \zo)^2 + (\partial_- z)^2\right] +2ig \sin z \partial_- z \psi_+\partial_- \psi_-.\end{eqnarray} By using the equations of motion the currents fulfil 
\begin{eqnarray} \label{conservi1} \partial_- T_4 &=& \partial_+ \theta_2 \\  \label{conservi2} \partial_+\overline{T}_4 &=& \partial_- \overline{\theta}_2, \end{eqnarray} and are therefore leading to conserved spin 3 quantities  in the bulk theory.

\section{$A(z),\;\overline{A}(\zo)$ - Identities}\label{aidentity}
Using the boundary conditions (\ref{bc1}) - (\ref{bc6}) we have the following identities at $\sigma=\pi$  used in the boundary expansion of the bulk conserved currents $T_4, \overline{T}_4$ and $\theta_2, \to_2$
 \begin{eqnarray} \partial_t \theta_- &=& \frac{1}{2}\pt \overline{A}(\zo)= \frac{1}{2} \overline{A}'(\zo)\;\partial_t \zo +\overline{F}' \overline{G}'\; \overline{\theta}_+ + \partial_z\partial_{\zo}B \;\theta_+\\ \partial_t \overline{\theta}_-  &=& \frac{1}{2} \pt A(z)=\frac{1}{2} A'(z)\; \partial_t z+\partial_z \partial_{\zo}B\; \overline{\theta}_+ + F' G' \; \theta_+ \end{eqnarray} \begin{eqnarray} \partial_t A'(z) &=& A''(z)\;\partial_t z + 2\; \partial_z\partial_z \partial_{\zo}B\; \overline{\theta}_++\partial_z\left(G'F'\right)\theta_+\\ \partial_t \overline{A}'(\zo)&=& \overline{A}''(\zo)\;\partial_t \zo+\partial_{\zo}\left(\overline{F}'\overline{G}'\right)\;\overline{\theta}_++2\;\partial_z\partial_{\zo}\partial_{\zo}B\;\theta_+\end{eqnarray} and
\begin{eqnarray} \partial_t A''(z) &=& A'''(z)\pt z + \left(G'''\overline{G}'+F'''\overline{F}'\right)\to_++\left(G'''F'+F'''G'\right)\theta_+\\ \partial_t \overline{A}''(\zo) &=& \overline{A}'''(\zo)\pt \zo+\left(\overline{F}'''\overline{G}'+\overline{G}'''\overline{F}'\right)\to_++\left(\overline{F}'''F'+\overline{G}'''G'\right)\theta_+.\end{eqnarray}
Quadratic fermionic terms as $A(z)\overline{A}(\zo)$ furthermore lead to identities like
\begin{eqnarray} A(z)\overline{A}(\zo) &=& \left(G'(z)\overline{G}'-F'(z)\overline{F}'(\zo)\right)\overline{a}a \label{aquadrat1}  \\ A(z)A'(z) &=& \left(\rule{0mm}{4mm}G'(z)F''(z)-F'(z)G''(z)\right)\overline{a}a.\label{aquadrat2}\end{eqnarray}  

\section{The boundary current $\Sigma_\pi^{(3)}(t)$}\label{sigmaidentity}
In this appendix we  present the explicit form of the local boundary term $\Sigma_\pi^{(3)}(t)$ appearing in the conserved  quantity  (\ref{spin3b}).  It is given by  
\begin{eqnarray} \nonumber  \Sigma_\pi^{(3)}(t) &=& 16i \pt^2 z A'(z)\theta_++16i\partial_t^2\zo \Ao'(\zo)\to_+
\\\nonumber && + 8\pz\pz B (\pt z)^2+8\pzo\pzo B (\pt\zo)^2+16\pz\pzo B\pt \zo \pt z  \\ \nonumber&& + 8ig^2\sin z \cos \zo\; A'(z) \theta_++8ig^2\sin \zo\; \cos z\; \Ao'(\zo)\; \to_+
\\ \nonumber && -4i \to_-\theta_+ ((\pz B)^2+3(\pzo B)^2)-4i\to_-\theta_+((\pt \zo)^2+3(\pt z)^2)\\ \nonumber &&
-8i\; \theta_-\to_+\; ((\pt \zo)^2+(\pz B)^2) \\ \nonumber && -16ig\cos z \eib\; \left(\theta_+\pt \theta_+-\theta_-\pt\theta_-\right) +32i \pt \to_-\pt\theta_+
\\ \nonumber && + 16ig^2\cos z \; \cos \zo\; \to_+\theta_- -8ig \;\sin\zo\; \eibb\; \pz B\; \to_+\to_-\\ \nonumber && +16i \to_+\; \overline{A}'(\zo) \;\pt^2\zo\; -16iA'(z)\pt^2 z \theta_+
 +32i \pz\pz\pzo B \pt z \to_+\theta_+ \\ \nonumber &&  +16i (\pt \zo)^2\; \to_+\overline{A}''(\zo) -16i (\pt z)^2 A''(z)\theta_+ 
\\ \nonumber &&  -8ig^2 \cos z \cos \zo\;\; \overline{\theta}_-\theta_+  +8ig \sin z\; \eib \pzo B\; \theta_-\theta_+ \\ \nonumber  && -8i \pz B \pt\zo\;\to_-\theta_- + 16 \pz B\; A'(z)\theta_-\to_+\;\theta_+  \\ && + H_1(z,\zo)+H_2(z,\zo) \end{eqnarray}
with 
\begin{eqnarray} \pz H_1(z,\zo)&=& \left(-2(\pz B)^3-6\pz B (\pzo B)^2\right) \\ \pzo H_1(z,\zo) &=&  \left(-2(\pzo B)^3-6\pzo B (\pz B)^2\right)\end{eqnarray}
 and 
\begin{eqnarray} \nonumber
\pz  H_2(z,\zo) &=&  +4g^2  \left[ \sin z\sin \zo\; \pzo B-\cos z \cos \zo \pz B\right. \\ &\;& \left. +2\sin z \cos \zo \pz\pz B +2\cos z \sin\zo\; \pz\pzo B\right]\\ 
\nonumber  \pzo H_2(z,\zo) &=& +4g^2  \left[\sin z \sin \zo \pz B-\cos z \cos \zo \; \pzo B +\right.\\ &\;& \left. + 2\sin z \cos \zo \pz\pzo B + 2\cos z \sin \zo \pzo\pzo B   \right].  \end{eqnarray}
For the choice $B(z,\zo) = \alpha \sin \frac{z}{2} \sin \frac{\zo}{2}$ these functions read
\begin{eqnarray} H_1(z,\zo) &=& \frac{1}{12} \alpha^2 \left(-4+\cos z+\cos\zo+2\cos z \cos\zo\right) B(z,\zo)\\ H_2(z,\zo)&=&  4g^2 B(z,\zo).\end{eqnarray}

\end{appendix}

\end{document}